\documentclass[conference,compsoc]{IEEEtran}
\usepackage[T1]{fontenc}
\usepackage{graphicx}
\usepackage{amsmath}
\usepackage{amssymb}
\usepackage{booktabs}
\usepackage{paralist}
\usepackage[breakable]{tcolorbox}
\usepackage{hyperref}
\usepackage{cite}

\usepackage{listings}
\newcommand{\toolname}{RAVEN}

\begin{document}

\title{\toolname: A Self-RAG Based Agent for Threat Classification and Response}

\author{
\IEEEauthorblockN{Francesco A. Pironti\textsuperscript{\dag}, Luigi Arena\textsuperscript{\dag}, Francesco Blefari\textsuperscript{\dag}, Matteo Lupinacci\textsuperscript{\dag}, Francesco Romeo\textsuperscript{\dag,*}, Angelo Furfaro\textsuperscript{\dag}}
    \IEEEauthorblockA{\textit{\dag DIMES - University of Calabria} \hspace{20pt} \textit{  *IMT School for Advanced Studies }\\
    Rende (CS) -- Italy  \hspace{28pt}  Lucca (LU) -- Italy \\
       {\{francesco.pironti, luigi.arena, francesco.blefari, matteo.lupinacci, francesco.romeo, angelo.furfaro\}@unical.it}}
    }

\maketitle

\begin{abstract}
Security event analysis in large enterprises can be overwhelming, as each alert typically requires manual inspection by human operators. This process is time-consuming and prone to inconsistencies, especially in high-volume Security Operations Center (SOC) environments. In this work-in-progress paper, we present \toolname{}: Retrieval-Augmented Vulnerability \& Event aNalysis, an agentic framework designed to support SOC analysts by automating the initial stages of alert investigation. The proposed system leverages a Self-Reflective Retrieval-Augmented Generation-based approach to provide an initial classification of security events and enables interactive refinement through analyst-in-the-loop queries. Unlike traditional approaches that rely on generic mitigation strategies, \toolname{} generates context-aware remediation suggestions by incorporating knowledge of the target enterprise infrastructure. We implement an early-stage prototype and report preliminary results on three balanced test sets of 400 curated web payloads each (SQLi, XSS, SSTI) and a qualitative case study of mitigation generation; an evaluation on IDS-generated alerts and a user study are ongoing and outlined as remaining work.
\end{abstract}


\section{Introduction}

In recent years, there has been a significant increase in cyberattacks targeting critical infrastructures and large-scale enterprises~\cite{ROUMANI2025104074}. Many of these organizations rely on SOCs~\cite{Vielberth2020SecurityOC} as their primary line of defense. These centers are inherently human-driven, with analysts responsible for determining whether observed events are benign or malicious. Given the growing complexity and volume of modern threats, it is essential to support SOC operators in making accurate and well-informed decisions.
While the adoption of Large Language Models~\cite{jaffal2025large} (LLMs) in combination with Retrieval-Augmented Generation~\cite{10.5555/3495724.3496517} (RAG) techniques has been increasingly explored in the cybersecurity domain~\cite{Ferrag20251}, recent studies have highlighted several limitations of RAG-based approaches, including issues related to retrieval quality, hallucinations, and contextual misalignment~\cite{weller2025theoretical}. To address these challenges, a novel paradigm known as Self-RAG has been proposed~\cite{asai2023selfraglearningretrievegenerate}. Self-RAG enhances the traditional RAG framework by enabling the model to autonomously decide when to retrieve external knowledge and to critically assess the relevance and correctness of the generated outputs. This capability is particularly relevant in cybersecurity contexts, where precision and reliability are crucial.
Despite these advancements, applying LLMs to SOC alert triage raises concrete challenges: (i) alert payloads are short and often ambiguous, so the model's parametric knowledge is unreliable for less common attack classes; (ii) generic mitigation advice is rarely applicable, since an effective countermeasure must target a specific device, host and service; (iii) alert data is sensitive and often cannot be sent to third-party APIs; and (iv) fine-tuning a classifier for each attack class is costly to maintain. \toolname{} addresses (i) with relevance-filtered retrieval, (ii) by grounding mitigations in an explicit infrastructure description, (iii) by running an open-weight model on-premises, and (iv) by being training-free.

In this work-in-progress paper we present \toolname{}, an agentic framework that sits downstream of an existing IDS/IPS and operates in two stages separated by an analyst checkpoint. In the first stage, a Self-RAG-inspired agent classifies the payload of each alert: it retrieves attack-specific documentation, discards retrieved context that an LLM judge deems irrelevant, reformulating the query when needed, and produces a structured verdict. Only after the analyst has approved this verdict does the second stage generate mitigations, namely a firewall rule for the perimeter device and configuration-hardening proposals, grounded in a lightweight YAML description of the target infrastructure. \toolname{} requires no fine-tuning: supporting a new attack category amounts to adding a knowledge base and extending the output schema. The work is guided by the following research questions:

\begin{itemize}
\item \textbf{RQ1:} Can a Self-RAG-based approach improve the accuracy and reliability of alert classification in SOC environments compared to traditional RAG-based methods?
\item \textbf{RQ2:} Can an agentic LLM workflow generate context-aware and actionable remediation strategies tailored to a specific enterprise infrastructure?
\end{itemize}

The main contributions of this work can be summarized as follows:
\begin{itemize}
\item An analyst-gated, two-stage pipeline coupling IDS/IPS alert triage with mitigation generation, so that no remediation stems from an unreviewed verdict.
\item A training-free, prompt-based Self-RAG instantiation in which an LLM judge filters retrieved context and drives query reformulation.
\item Mitigation generation grounded in a lightweight YAML infrastructure description, yielding device-level artifacts (a perimeter firewall rule and host-specific hardening proposals).
\item  An on-premises prototype based on an open-weight model and a   preliminary ablation on three balanced test sets (SQLi, XSS, SSTI),   showing that retrieval helps mostly on context-dependent attacks (SSTI).
\end{itemize}
The paper is organized as follows: Section~\ref{sec:background} presents background and related work, Section~\ref{sec:methodology} describes \toolname{}, Section~\ref{sec:evaluation} reports the evaluation, Section~\ref{sec:discussion} discusses results and remaining work, and Section~\ref{sec:conclusion} concludes.

\section{Background and Related Work}
\label{sec:background}

This section introduces the background notions on which \toolname{} is built and the related work against which it is positioned. Among the many forms of cyber-attack, web-based attacks are particularly dangerous because of their low barrier to entry: they allow an unprivileged and often anonymous actor, such as a casual visitor to a public-facing website, to gain an initial foothold within an otherwise well-defended corporate network. This combination of wide exposure and comparatively simple exploitation makes them both a frequent occurrence in real SOC alert queues and a natural starting point for evaluating an automated triage system. We focus on three representative and well-studied vectors.
\textit{Cross-Site Scripting (XSS)} exploits JavaScript code injection~\cite{gupta2017,HYDARA2015}, enabling an unauthenticated external entity to execute injected code within the victim's web browser. \textit{Server-Side Template Injection (SSTI)} arises when an attacker can inject malicious payloads into a server-side template, exploiting the template engine to execute unauthorized code on the server~\cite{Zhao2023}. \textit{SQL Injection} occurs when an attacker can insert and execute malicious code within a vulnerable SQL query~\cite{Singh2016}.

\subsection{Security Operations Centers and IDS/IPS}

A Security Operations Center~\cite{Vielberth2020SecurityOC} is a centralized function that continuously monitors, assesses, and defends an enterprise's information systems. The primary role of a SOC is to enhance the institutional security posture through the real-time detection and mitigation of cyber threats. To this end, it integrates advanced telemetry such as Security Information and Event Management (SIEM) systems and Endpoint Detection and Response (EDR) tools. An Intrusion Detection System (IDS) monitors the traffic of the protected infrastructure and raises an alert when it observes a threat, either by matching traffic against signatures of known attacks (e.g., specific keywords in HTTP requests) or by detecting deviations from a learned model of normal behavior~\cite{debar2009}; security operators are then responsible for deciding the appropriate response. An Intrusion Prevention System (IPS) is positioned between protected hosts and external networks and can also act on the traffic it inspects, for instance by blocking a connection.

\subsection{AI Agents, Agentic RAG and Self-RAG}

An agent is a computational system situated within an environment that can act autonomously to achieve its delegated objectives~\cite{wooldridge:2009}. 
An intelligent agent is characterized by its ability to perceive its environment, respond to changes within it, initiate actions, and interact with other systems including other agents or human users. One of the core principles foundational to AI agents is the concept of memory: effective memory management enhances an agent's ability to maintain contextual awareness, enabling it to draw on previous experiences and develop progressively more informed decision-making over time. 
LLM agents extend the capabilities of LLMs by enabling the use of external tools for specialized tasks, including mathematical computation and code execution; the model continuously evaluates the adequacy of its output and determines whether additional processing steps are required. 
General-purpose LLMs often lack the domain-specific knowledge required to solve specialized problems. Fine-tuning can address this limitation but is typically costly and resource-intensive. RAG~\cite{10.5555/3495724.3496517} offers a practical alternative by augmenting LLMs with external knowledge sources accessed through a retriever component. In a standard RAG pipeline, external documents are segmented, embedded, and stored for similarity-based retrieval. The workflow consists of (i)~submitting a user query to a RAG agent; (ii)~retrieving the most relevant document chunks via vector similarity; (iii)~augmenting the query with the retrieved context; (iv)~generating a context-aware response using the LLM. Agentic RAG systems integrate this paradigm within an agent framework, enabling iterative reasoning and tool use.

\begin{figure}[t]
\centering
\includegraphics[width=\columnwidth]{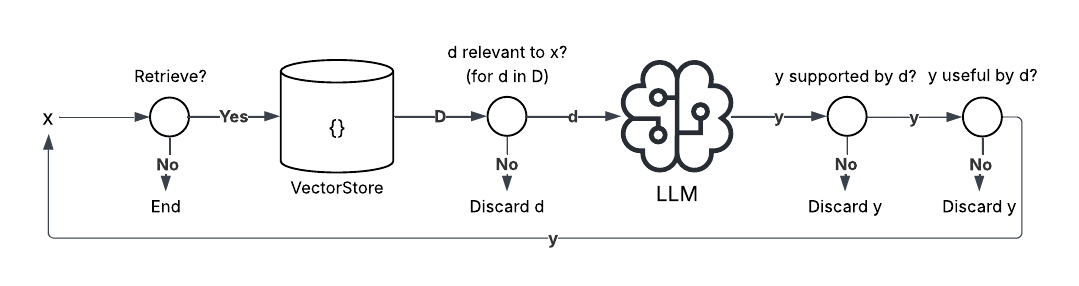}
\caption{Generic Self-RAG workflow~\cite{asai2023selfraglearningretrievegenerate}. Given an input query $x$, the system retrieves a set of candidate documents $D$, which are filtered based on relevance. A candidate response $y$ is then generated by an LLM and evaluated through a sequential self-reflection procedure defined over documents $d \in D$.}
\label{fig:self-rag}
\end{figure}

Self-RAG~\cite{asai2023selfraglearningretrievegenerate} extends the RAG paradigm by introducing reflection-based validation steps to improve retrieval and generation quality. As shown in Fig.~\ref{fig:self-rag}, given an input query $x$, the system retrieves a set of candidate documents $D$, which are filtered based on relevance. A candidate response $y$ is then generated by an LLM and evaluated through a sequential self-reflection procedure defined over documents $d \in D$, based on the following queries:
\begin{enumerate}
\item \textbf{SQ1:} Is $y$ adequately supported by a retrieved document $d \in D$?
\item \textbf{SQ2:} Is $y$ informative and pertinent with respect to $x$?
\end{enumerate}
In the original formulation~\cite{asai2023selfraglearningretrievegenerate}, these decisions are expressed through reflection tokens emitted by a model trained for this purpose. \toolname{} adopts a lightweight, prompt-based instantiation of the paradigm that requires no training: the relevance critique over retrieved documents is delegated to a separate LLM call acting as a judge, which also drives query reformulation, while the generated output is checked only for structural validity (SQ1 and SQ2 are not implemented in the current prototype).

\subsection{LLM-Based Alert Triage and Incident Response}
\label{sec:related_work}

Building on the paradigms above, we group the most closely related LLM-based efforts into alert triage and incident response, and then position \toolname{} with respect to them.

\paragraph{\textit{Alert triage}} CyberRAG~\cite{cyberrag} is an agent that takes input from an IDS/IPS and, using BERT-based classifiers, selects an appropriate knowledge base, performs a query, and uses the findings to generate a report; however, its classifiers must be fine-tuned for each new attack category, and it lacks an explicit focus on mitigation. Sch\"armer et al.~\cite{scharmer2026alerts} evaluate commercial LLMs (ChatGPT and Gemini) on the classification of alerts produced by network- and host-based IDS, and study how system context, such as additional logs, configuration data, and few-shot examples, affects classification performance and consistency; they also point out limitations related to cost, reliability, and the exposure of sensitive data. Their focus is on interpreting alerts rather than on proposing remediations. The approach in~\cite{Hmimou2025AMS} leverages multi-agent architectures combined with LLMs to enhance threat detection across heterogeneous data sources, achieving improved accuracy and cross-domain correlation; however, it primarily focuses on detection and analysis, without support for infrastructure-grounded response.

\paragraph{\textit{Incident response}} The authors in~\cite{rajapaksha2024rag} introduced a RAG-based Question-Answering model designed to assist experts in cyber-attack investigation and attribution. By leveraging a curated knowledge base, the model outperforms general-purpose LLMs like GPT-4o in providing source-backed responses and reducing hallucinations; however, its scope is informational and reactive, without autonomous attack analysis or remediation suggestions. Hammar et al.~\cite{hammar2026incident} combine instruction fine-tuning of a lightweight LLM, retrieval of threat information, and decision-theoretic lookahead planning to generate incident-response plans with a bounded probability of hallucination, evaluated on logs from incidents reported in the literature. IRCopilot~\cite{lin2025ircopilot} mimics the phases of a real-world incident response team through four collaborating LLM components (Planner, Generator, Reflector, and Analyst), and is evaluated on a benchmark built from incident-response challenges on public platforms. Both works address response planning for an incident that is already under way, rather than the triage of individual IDS alerts.

\paragraph{\textit{Positioning}} \toolname{} differs from these works in three respects. First, it targets the earliest stage of the SOC workflow, the triage of individual IDS/IPS alerts, and couples it with mitigation in a single pipeline, separated by an explicit analyst checkpoint so that no remediation is generated from an unreviewed verdict. Second, it requires no fine-tuning, unlike~\cite{cyberrag,hammar2026incident}: domain knowledge is supplied through per-attack knowledge bases whose retrieved content is filtered by a relevance judge, and the prototype relies on an open-weight model run on-premises, which avoids sending alert data to third-party services, a concern raised in~\cite{scharmer2026alerts}. Third, its mitigations take the form of concrete, device-level artifacts, namely a rule for the perimeter device and configuration changes, that reference host identifiers, services, and ports from an explicit, lightweight infrastructure description. Unlike~\cite{scharmer2026alerts}, our current evaluation does not use alerts produced by a running IDS, and unlike~\cite{hammar2026incident,lin2025ircopilot}, it does not yet evaluate response quality quantitatively.

\section{\toolname}
\label{sec:methodology}

\toolname{} is an agentic workflow designed to support SOC operations by triaging alerts generated by an existing IDS/IPS, such as Suricata~\cite{suricataids} or Snort~\cite{roesch1999snort}, whose output it refines. These systems record each alert as a JSON entry (for Suricata, the EVE JSON log) that contains both the network information, such as IP addresses and ports, and the payload. For each alert, \toolname{} produces a preliminary report that summarizes the potential threat, grounded both in the model's parametric knowledge and in the documentation retrieved from its knowledge base (RAG). Once the analyst has reviewed and approved the report, the system attempts to generate actionable instructions to mitigate the threat, leveraging a textual description of the target enterprise infrastructure so that recommendations are context-aware and tailored to the specific environment.

\begin{figure*}[t]
\centering
\includegraphics[width=1.6\columnwidth]{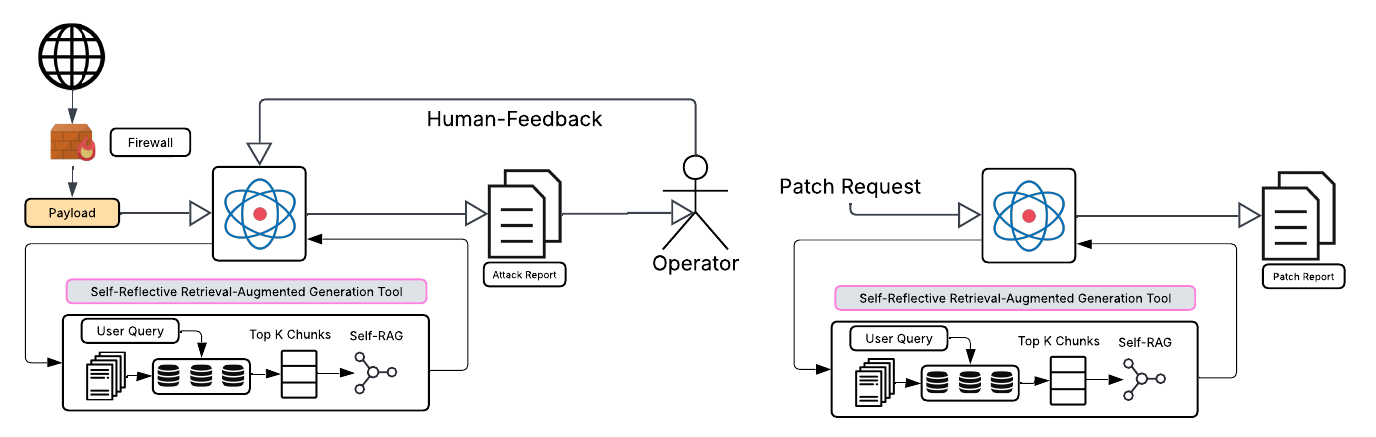}
\caption{\toolname{} architecture.}
\label{fig:architecture}
\end{figure*}

The proposed system is orchestrated as a stateful, directed graph using the LangGraph framework~\cite{langgraph} and is divided into specialized nodes, each responsible for a discrete phase of the analysis pipeline (Fig.~\ref{fig:architecture}). The pipeline comprises two parts. The first is an agentic workflow that needs access only to the alert payload and produces a structured report as output. The analyst reviews the report and may then proceed to the second part, which takes the structured report as input, together with the full JSON alert record, and proposes a mitigation tailored to the target infrastructure.

\subsection{Lightweight YAML-Based Infrastructure Representation}

Formal ontologies such as the Unified Cybersecurity Ontology (UCO)~\cite{Syed2016} offer high semantic expressiveness and interoperability, but rely on RDF/OWL representations (e.g., JSON-LD) that require graph-based parsing, ontology reasoning, and non-trivial data transformation pipelines. Since our objective is not full semantic interoperability but an infrastructure representation an LLM can reason over efficiently, we adopt a simplified YAML-based representation that trades formal rigor for usability, readability, and ease of processing. The schema is intentionally minimal and structured around four components:

\begin{itemize}
\item \textbf{hosts}: the entities in the infrastructure, each identified by an \texttt{id}, an \texttt{ip} address, and a \texttt{role} (e.g., firewall, web server, router).
\item \textbf{services}: network services associated with hosts, specifying the service \texttt{name} and \texttt{port}, which makes the attack surface explicit.
\item \textbf{links}: connectivity and logical relationships between hosts, each defined by a \texttt{from} and \texttt{to} field, with an optional \texttt{note}.
\item \textbf{vulns}: known vulnerabilities associated with hosts, using lightweight identifiers (e.g., CVE IDs); this component is optional and is not used in our testbed.
\end{itemize}

This flat and declarative structure avoids deep nesting and complex semantics, making it particularly suitable for LLM-based processing.

\subsection{Attack Detection and Reporting}
\label{sec:detection}

\paragraph{\textit{Query Generator Node}} The pipeline starts by using an LLM to extract vulnerability syntax patterns from the raw payload and to reformulate them into a concise, technically precise search query optimized for vector similarity search, rather than using the raw payload for retrieval.

\paragraph{\textit{Retrieve Node}} The generated query is used to fetch the top-$k$ most relevant chunks from the vector store, which are aggregated into a context string with their source metadata to ground generation in verifiable documentation.

\paragraph{\textit{Relevance Node}} An LLM judge assesses whether each retrieved chunk contains actionable information related to the payload's potential attack techniques. Chunks deemed \texttt{NOT\_RELEVANT} are discarded and the graph routes back to the Query Generator, which reformulates the query to retrieve replacements, up to a fixed number of iterations; once all chunks are deemed \texttt{RELEVANT}, or the limit is reached, the pipeline advances to generation.

\paragraph{\textit{Generate Node}} The Generate Node serves as the core analytical engine of the architecture. It ingests the original payload together with the retrieved chunks that passed the relevance check. The LLM is prompted to act as a cybersecurity analyst, evaluating the payload for malicious intent, obfuscation techniques, and vulnerability probes, and its output is constrained to a predefined JSON schema.

\paragraph{\textit{Validation and Repair Nodes}} To guarantee downstream interoperability, a \textbf{JSON Check Node} verifies that the output is well-formed and contains all required keys, terminating the graph on success, while a \textbf{JSON Fix Node} prompts the LLM to repair malformed output within a retry limit, falling back to full regeneration or an error state.

\paragraph{\textit{Structured Output Schema}} The final report is a JSON object with the fields \texttt{verdict} (\texttt{BENIGN}/\texttt{MALICIOUS}), \texttt{attack\_type} (\texttt{SQLi}, \texttt{XSS}, \texttt{SSTI}, \texttt{OTHER}, or \texttt{NONE}), \texttt{risk\_level} (\texttt{LOW}/\texttt{MEDIUM}/\texttt{HIGH}), \texttt{confidence} $\in [0,1]$, and a human-readable \texttt{technical\_explanation} (prompts in Appendix~\ref{appendix:prompts}).

\subsection{Mitigation Proposal}

After the structured report has been approved by the analyst, the mitigation stage takes as input the report and the full EVE JSON alert. It first checks whether the targeted host is exposed to the detected attack; if not, an explanation is returned to the analyst; otherwise, two types of mitigation grounded in the infrastructure description are proposed: a firewall rule to be applied manually, and configuration changes.

\paragraph{\textit{Vulnerability check Node}}
Given the infrastructure YAML, the EVE
alert and the verdict, the LLM identifies the host and service targeted by the alert and inspects their configuration and software versions, retrieved from the service configurations stored in the knowledge base (Section 4), to determine whether that version and
configuration could actually be affected by the detected attack (e.g., a template engine that renders user input for SSTI). If the service is deemed not vulnerable, the graph terminates with an explanation for the analyst; otherwise it proceeds to the Firewall
retrieve Node.

\paragraph{\textit{Firewall retrieve Node}} This node retrieves firewall-rule patterns from an external knowledge base. Given the raw EVE JSON alert and the structured verdict, it prompts the LLM to synthesize a concise retrieval query (at most ten words) targeting mitigation-via-firewalling documentation. The query is forwarded to the RAG engine, which returns the top-$k$ most relevant chunks, stored in the \texttt{fw\_context} field of the shared state.

\paragraph{\textit{Firewall generate Node}} This node synthesizes a concrete firewall rule targeting the network boundary device. The context available at this stage comprises (i)~the verbatim infrastructure YAML, (ii)~the raw EVE alert, (iii)~the trusted analyzer verdict, and (iv)~the retrieved firewall patterns. The model identifies the host whose declared role corresponds to a perimeter device (\texttt{firewall}, \texttt{gateway}, \texttt{router}) and writes the rule for that host; in the absence of such a host, it declares this condition explicitly and produces a generic fallback rule.

\paragraph{\textit{Hardening retrieve Node}} This node mirrors the retrieval logic of the firewall-retrieve node but targets a different subdomain of the knowledge base, generating a query oriented towards hardening guides, configuration best practices, and security benchmarks relevant to the observed attack category; results are stored in \texttt{hardening\_context}.

\paragraph{\textit{Proposals generate Node}} This node receives the complete accumulated state (infrastructure descriptor, EVE alert, analyzer verdict, the firewall rule produced by the previous node, and the hardening context) and produces a structured natural-language report of configuration-change proposals. Every recommendation references specific host identifiers, service names, and port numbers declared in the infrastructure YAML, paired with a security rationale, ensuring traceability between the observed threat and the recommended countermeasure.

\section{Evaluation}
\label{sec:evaluation}

For this evaluation we constructed three distinct knowledge bases, one for each attack type under investigation (SSTI, SQL Injection, XSS). Documentation for the threat-reporting knowledge base was sourced from PortSwigger Academy~\cite{portswigger_academy} and OWASP guides~\cite{owasp_top10_2025}; for the mitigation knowledge bases, one source is a firewall-rule-writing guide, and the other is a copy of the actual service configurations and specific service guides.
Each knowledge base was split into chunks of 500 characters with an overlap of 150 characters to preserve context across boundaries, and each chunk was transformed into embeddings using a pre-trained Sentence Transformer model, Qwen3-Embedding~\cite{zhang2025qwen3}. Embeddings were stored in in-memory vector stores using the Faiss library~\cite{douze2025faisslibrary}, paired with their corresponding chunk, using Cosine Similarity as the similarity metric.

\subsection{Attack Classification and Ablation Study}

The attack dataset, made publicly available in~\cite{cyberrag}, contains three attack classes (SQL injection, SSTI, and XSS) and a benign class; it was chosen for the availability of both attack and normal-operation samples. For each attack type we built a balanced test set by sampling 200 malicious payloads of that type and 200 benign payloads from the
larger pools of the dataset, yielding three test sets of 400 samples each. Metrics refer to binary malicious/benign classification (the \texttt{verdict} field) on each test set; attack-type identification is not reflected in these figures. These samples are payloads from a curated dataset rather than alerts raised by a running IDS, a limitation we discuss in Section~\ref{sec:discussion}.
To assess the contribution of retrieval and of the Self-RAG paradigm, we compare three configurations sharing the same LLM (Qwen3:32B~\cite{qwen3}), prompt structure, and output schema: \textbf{LLM-only}, without external knowledge; \textbf{RAG}, where retrieved context is always provided; and \textbf{Self-RAG}, the full system, where retrieved context is filtered by the relevance judge and the query is reformulated when needed (Section~\ref{sec:detection}). Results are reported in Table~\ref{tab:ablation}.

\begin{table}[t]
\centering
\small
\setlength{\tabcolsep}{4pt}
\caption{Ablation results per attack type (best value per attack in bold; ties not marked).}
\label{tab:ablation}
\begin{tabular}{llcccc}
\toprule
\textbf{Attack} & \textbf{Config.} & \textbf{Acc.} & \textbf{Prec.} & \textbf{Rec.} & \textbf{F1} \\
\midrule
SQLi & LLM-only          & 0.96 & 0.94 & 0.98 & 0.96 \\
     & RAG               & 0.98 & 0.97 & 0.98 & 0.97 \\
     & \textbf{Self-RAG} & \textbf{0.99} & \textbf{1.00} & 0.98 & \textbf{0.99} \\
\midrule
XSS  & LLM-only          & 0.99 & \textbf{1.00} & 0.98 & 0.99 \\
     & RAG               & 0.99 & 0.99 & 0.99 & 0.99 \\
     & \textbf{Self-RAG} & 0.99 & 0.98 & \textbf{1.00} & 0.99 \\
\midrule
SSTI & LLM-only          & 0.84 & 0.84 & 0.84 & 0.84 \\
     & RAG               & 0.89 & 0.88 & 0.90 & 0.89 \\
     & \textbf{Self-RAG} & \textbf{0.92} & \textbf{0.90} & \textbf{0.94} & \textbf{0.92} \\
\bottomrule
\end{tabular}
\end{table}

The full system reaches an F1 of 0.99 on SQLi and XSS and 0.92 on SSTI. On SQLi and XSS, which exhibit well-defined and recognizable patterns, the LLM-only baseline is already close to ceiling: retrieval raises the F1 on SQLi from 0.96 to 0.99, mainly by removing false positives, while on XSS all three configurations reach 0.99 and differ only in the balance between precision and recall. The benefit is concentrated on SSTI, whose payloads are more variable and dependent on application context: F1 rises from 0.84 (LLM-only) to 0.89 (RAG) and 0.92 (Self-RAG). External knowledge thus matters most where the model's internal knowledge of attack patterns is least reliable, and filtering irrelevant context brings a further, smaller gain over always providing it. The residual false negatives on SSTI indicate that further improvements are needed where missing an attack may have critical consequences; moreover, since results come from a single run per configuration, differences of one or two points should be treated with caution.

\subsection{Mitigation Generation}

\paragraph{\textit{Testbed Configuration}} Since evaluating generated remediation is inherently challenging, we use a controlled and reproducible setup: a minimal test infrastructure simulating a typical web application scenario, consisting of (i)~an external router, (ii)~a Suricata-based firewall operating in IPS mode, and (iii)~a web server hosting a vulnerable FastAPI/Jinja2 application, described using the YAML representation of Section~\ref{sec:methodology}:

\begin{tcolorbox}[left=1mm, right=1mm, top=0.5mm, bottom=0.5mm, boxsep=0.5mm]
\scriptsize
\begin{lstlisting}
hosts:
- {id: router, ip: 203.0.113.1, role: external_router}
- {id: firewall, ip: 192.168.1.1, role: firewall_ips}
- {id: webserver, ip: 192.168.1.100, role: web_server}
services:
- {host: webserver, name: http server, port: 80,
   note: "Jinja2, FastAPI"}
- {host: firewall, name: suricata, port: 0,
   note: "IPS mode, inspects traffic"}
links:
- {from: router, to: firewall,
   note: external traffic flows through firewall}
- {from: firewall, to: webserver, note: filtered HTTP traffic}
\end{lstlisting}
\end{tcolorbox}

\paragraph{\textit{Evaluation Methodology}} We adopt a lightweight qualitative assessment based on three criteria: \textbf{Actionability} (whether proposed mitigations could be directly applied or translated into concrete configuration changes); \textbf{Correctness} (whether the mitigation effectively addresses the identified threat without introducing inconsistencies or unsafe configurations); and \textbf{Context-awareness} (whether the mitigation leverages the provided infrastructure description rather than proposing generic strategies).

\paragraph{\textit{Generated Mitigations}} In scenarios involving network-level attacks, the system generates firewalling rules tailored to the infrastructure. For example, in the presence of malicious HTTP payloads targeting the web server, the model proposes rules that restrict or filter traffic based on source, destination, and port---correctly identifying the \texttt{firewall\_ips}-role host and referencing concrete host, service, and port identifiers from the YAML model in its hardening proposals.

\section{Discussion \& Remaining Work}
\label{sec:discussion}
The ablation study highlights three key observations. First, \textit{impact of retrieval}: RAG and Self-RAG outperform the LLM-only baseline on SQLi and SSTI and tie with it on XSS, so external knowledge matters most where the model's internal knowledge is weaker. Second, \textit{limitations of standard RAG}: always providing retrieved context may include chunks unrelated to the payload; this noise may explain part of the gap to Self-RAG, although our experiments do not isolate it. Third, \textit{relevance filtering}: discarding irrelevant chunks and reformulating the query yields a further, smaller gain, most visible on SSTI. Regarding \textbf{RQ1}, the ablation provides preliminary and partial support: retrieval and relevance filtering help most on context-dependent attacks such as SSTI, whereas on well-patterned attacks simpler configurations already perform comparably, and the advantage of Self-RAG over standard RAG is modest (0--3 F1 points). Regarding \textbf{RQ2}, the qualitative case study suggests that grounding generation in a lightweight, structured description of the target infrastructure is enough to move proposals beyond generic security advice: rules and hardening suggestions are anchored to actual host roles, service names, and ports rather than to abstract attack categories. The bottleneck for actionable remediation thus appears to be less the sophistication of the infrastructure model than its availability and accuracy, since any staleness or omission in the YAML description propagates directly into an incorrect or incomplete recommendation.
These results are preliminary, and several aspects of \toolname{} require further investigation:

\begin{itemize}
\item \textbf{Realistic alert triage.} Evaluating \toolname{} on alerts raised by an IDS, e.g., by replaying the dataset through the Suricata testbed, so as to include benign traffic that triggers signatures, which is harder to separate from attacks; attack-type identification will also be reported separately.
\item \textbf{Baselines and cost.} Comparing Self-RAG with optimized prompt-only baselines (e.g., few-shot prompting) and, quantitatively, with CyberRAG~\cite{cyberrag}, and reporting latency and LLM calls per alert, to establish whether the gains justify the added complexity.
\item \textbf{Statistical robustness.} Repeating each configuration across random seeds and sampling temperatures to assess the variance of both classification and mitigation outputs.
\item \textbf{Quantitative mitigation assessment.} Applying the generated rules in the testbed and replaying malicious and legitimate traffic to measure blocked attacks and service disruption, also assessing the contribution of the YAML description and the effect of incorrect verdicts on the proposals.
\item \textbf{Human-subject evaluation and coverage.} A user study with SOC analysts to quantify reductions in cognitive load and improvements in Mean Time To Detect/Respond (MTTD/MTTR), and an extension to further attack classes (e.g., command injection, deserialization). The analyst checkpoint is currently a design safeguard rather than an evaluated component: its benefit, and the cost of an incorrect verdict reaching the mitigation stage, will be measured in the planned user study and testbed experiments.

\item{ \textbf{Confidence calibration and alert prioritization.}} The \texttt{confidence} and \texttt{risk\_level} fields of the structured report are currently not evaluated. We plan to assess their calibration against ground truth and to use them to rank alerts in the analyst's queue, so that low-confidence verdicts are flagged for closer inspection.
\end{itemize}

\section{Conclusion}
\label{sec:conclusion}

In this work-in-progress paper we presented \toolname{}, an analyst-gated agentic framework that couples IDS/IPS alert triage, based on a training-free Self-RAG-inspired workflow, with mitigation generation grounded in a lightweight description of the target infrastructure. A preliminary ablation on three balanced test sets of 400 payloads
each shows that retrieval helps mostly on context-dependent attacks such as SSTI (F1 from 0.84 to 0.92), while on well-patterned attacks an LLM-only baseline is already close to ceiling. A qualitative case study suggests that infrastructure grounding yields device-level, context-aware proposals. The evaluation plan outlined in Section~\ref{sec:discussion}, from IDS-generated alerts to a user study with SOC analysts, will establish whether these gains hold in realistic triage.

\section*{Acknowledgment}
This work has been partially supported by Olivgrow project (CUP E29I26001100005), funded by "Azione 1.1.1 PR Calabria FESR 2021-2027”.
This work has been partially supported within the technology transfer activities of the Second-Level Master’s Program in Electric Mobility and Circular Economy framework, funded by the Italian MUR under the “Territorial Pacts for Higher Education for Enterprises” (CUP H52C23000090001).
The work of F.A. Pironti, M. Lupinacci and L. Arena was supported by Agenzia per la cybersicurezza nazionale under the program for the promotion of XL,XLI,XLII cycle PhD research in cybersecurity – (CUP H23C24000640005, H23C25000360005, H23C26000400005). The views expressed are those of the authors and do not represent the funding institution.

\appendix
\tcbset{left=1mm, right=1mm, top=0.5mm, bottom=0.5mm, boxsep=0.5mm, breakable}

\section{Threat Reporting Prompts}
\label{appendix:prompts}

\begin{tcolorbox}[title=Query Generator Node]
\scriptsize
\begin{lstlisting}
You are refining a search query for a cybersecurity knowledge base.

INPUT PAYLOAD:
{state['payload'][:800]}

Task:
Generate a short search query to retrieve documentation about 
similar attack payloads.
Rules:
- Focus on vulnerability syntax patterns present in the payload
- Use precise security terminology (e.g. "sql injection", "xss",
"ssti", "path traversal")
- Avoid natural language sentences
- Maximum 10 words
Examples:
  Payload: {{7*7}} -> Query: ssti template 
        injection arithmetic probe
  Payload: <script>alert(1)</script> ->
  Query: xss reflected script injection payload
  Payload: ' OR 1=1,  ->
  Query: sql injection authentication bypass union
Return ONE LINE ONLY: the query, nothing else.
\end{lstlisting}
\end{tcolorbox}

\begin{tcolorbox}[title=Relevance Node (Self-RAG)]
\scriptsize
\begin{lstlisting}
You are a cybersecurity expert evaluating retrieved context.

INPUT PAYLOAD:
{state['payload'][:300]}

RETRIEVED CONTEXT:
{state['context']}

Task:
Decide whether the retrieved context contains information that would help analyze
the payload as a potential attack.

Decision rules:
- Context does NOT need to match the payload verbatim; related attack techniques count.
- Answer NOT_RELEVANT if the payload is:
    - plain text, a common word, a standalone number
    - a simple HTML structural tag (<ul>, <div>, <p>, </form>, etc.)
    - a random string, encoded token, or noisy fragment with no attack syntax
- Answer RELEVANT if the context discusses techniques, patterns, or vectors that
  resemble what is present in the payload.

Answer with exactly ONE word: RELEVANT or NOT_RELEVANT
\end{lstlisting}
\end{tcolorbox}

\begin{tcolorbox}[title=Generate Node]
\scriptsize
\begin{lstlisting}
You are a cybersecurity analyst performing threat detection.

Input payload:
{payload_excerpt}
{context_section}

Task:
Analyze the payload and determine whether it is benign or malicious.
Consider the retrieved context if provided.

Return STRICT JSON with exactly these fields:
{
  "verdict": "BENIGN | MALICIOUS",
  "attack_type": "NONE | SQLi | XSS | SSTI | OTHER",
  "confidence": <float 0.0-1.0>,
  "technical_explanation": "<detailed explanation>",
  "risk_level": "LOW | MEDIUM | HIGH"
}

Rules:
- BENIGN -> attack_type must be NONE, risk_level must be LOW
- MALICIOUS -> risk_level must be MEDIUM or HIGH
- A payload that probes for vulnerabilities is MALICIOUS even if it does not 
  fully exploit them.
- A standalone SQL SELECT statement provided as plain input is BENIGN.
- Obfuscation techniques (base64, hex encoding, template syntax) are indicators 
  of malicious intent.
- Focus on the presence of "eval", "base64_decode", "__subclasses__", or similar 
  dangerous patterns rather than attempting to manually decode the full payload.
- Output ONLY valid JSON: no preamble, no markdown fences.
\end{lstlisting}
\end{tcolorbox}

\section{Mitigation Generation Prompts}

The Firewall and Hardening retrieve nodes use a query-generation prompt analogous to the Query Generator, asking for a query of at most ten words given the EVE alert and the analyzer verdict. The two generation prompts are reported below.
\begin{tcolorbox}[colback=gray!8, colframe=black!75,
  title=Vulnerability Check Node, fonttitle=\small,
  boxrule=0.8pt, arc=2mm]
\begin{lstlisting}[basicstyle=\ttfamily\scriptsize,
  breaklines=true, columns=fullflexible, upquote=true]
You are a vulnerability assessment engineer.
INFRASTRUCTURE:
{state['infra_raw']}
EVE ALERT:
{state['eve_raw']}
PAYLOAD ANALYSIS (trusted):
{json.dumps(state['analyzer_result'], indent=2)}
SERVICE CONFIGURATIONS:
{state['config_context']}
Identify the targeted service and check whether its
configuration and versions could be affected by this
attack. Answer VULNERABLE or NOT_VULNERABLE on the
first line, followed by a brief justification.
\end{lstlisting}
\end{tcolorbox}

\begin{tcolorbox}[title=Firewall Generate Node]
\scriptsize
\begin{lstlisting}
You are a network security engineer.
INFRASTRUCTURE:
{state['infra_raw']}
EVE ALERT:
{state['eve_raw']}
PAYLOAD ANALYSIS (trusted):
{json.dumps(state['analyzer_result'], indent=2)}
RETRIEVED FIREWALL PATTERNS:
{state['fw_context']}
Generate a concrete firewalling rule to block or mitigate this attack.
Include a brief justification after the rule. Check for errors.
\end{lstlisting}
\end{tcolorbox}

\begin{tcolorbox}[title=Proposals Generate Node]
\scriptsize
\begin{lstlisting}
You are a senior infrastructure security architect.
INFRASTRUCTURE:
{state['infra_raw']}
EVE ALERT:
{state['eve_raw']}
PAYLOAD ANALYSIS (trusted):
{json.dumps(state['analyzer_result'], indent=2)}
FIREWALL RULE (already proposed):
{state['fw_rule']}
RETRIEVED HARDENING CONTEXT:
{state['hardening_context']}
Propose specific configuration changes to harden the infrastructure
against this attack. Reference host IDs, ports and services
from the infrastructure description where applicable.
For each change, state what to modify and why.
\end{lstlisting}
\end{tcolorbox}

\bibliographystyle{IEEEtran}
\bibliography{ref}

\end{document}